\documentclass{aa}

\usepackage{graphicx}
\usepackage{txfonts}
\usepackage{hyperref}
\usepackage{booktabs}
\usepackage{cleveref}
\usepackage{etoolbox}
\usepackage[all]{hypcap}
\usepackage{orcidlink}
\usepackage{upgreek}

\makeatletter
\renewcommand*\aa@pageof{, page \thepage{} of \pageref*{LastPage}}
\makeatother

\hypersetup{
    hidelinks,
    colorlinks=true,
    linkcolor=blue,
    urlcolor=blue,
    citecolor=blue,
}

\crefname{equation}{Eq.}{Eqs.}
\Crefname{equation}{Equation}{Equations}
\crefname{figure}{Fig.}{Figs.}
\Crefname{figure}{Figure}{Figures}
\crefname{table}{Table}{Tables}
\Crefname{table}{Table}{Tables}
\crefname{section}{Sect.}{Sects.}
\Crefname{section}{Section}{Sections}

\makeatletter
\renewcommand*\aa@authorrunning{Robert Lilow et al.}
\makeatother

\defcitealias{ganeshaiah_veena_large-scale_2023}{Paper I}
\defcitealias{lilow_constrained_2021}{LN21}

\begin{document} 

\title{Neural network reconstruction of density and velocity fields from the 2MASS Redshift Survey}
\titlerunning{Neural network reconstruction of density and velocity fields from 2MRS}

\author{Robert Lilow\inst{\ref{inst1}}\inst{\ref{inst2}}\thanks{Corresponding author}\orcidlink{0000-0002-6960-6065}
\and Punyakoti Ganeshaiah Veena\inst{\ref{inst2}}\orcidlink{0000-0003-4053-1749}
\and Adi Nusser\inst{\ref{inst2}}\orcidlink{0000-0002-8272-4779}
}

\institute{Institute for Theoretical Physics, Heidelberg University, 69120 Heidelberg, Germany
\label{inst1}\\
\email{lilow@thphys.uni-heidelberg.de}
\and
Department of Physics, Technion, Haifa 3200003, Israel
\label{inst2}\\
\email{punyakoti.g@campus.technion.ac.il}\\
\email{adi@physics.technion.ac.il}
}

\date{Received 2 April 2024 /
Accepted 30 May 2024}

\abstract
{}
{Our aim is to reconstruct the 3D matter density and peculiar velocity fields in the local Universe up to a distance of $200\,h^{-1}\,\mathrm{Mpc}$ from the Two-Micron All-Sky Redshift Survey (2MRS) using a neural network (NN).}
{We employed an NN with a U-net autoencoder architecture and a weighted mean squared error loss function trained separately to output either the density or velocity field for a given input grid of galaxy number counts. The NN was trained on mocks derived from the Quijote N-body simulations, incorporating redshift-space distortions (RSDs), galaxy bias, and selection effects closely mimicking the characteristics of 2MRS. The trained NN was benchmarked against a standard Wiener filter (WF) on a validation set of mocks before applying it to 2MRS.}
{The NN reconstructions effectively approximate the mean posterior estimate of the true density and velocity fields conditioned on the observations. They consistently outperform the WF in terms of reconstruction accuracy and effectively capture the nonlinear relation between velocity and density. The NN-reconstructed bulk flow of the total survey volume exhibits a significant correlation with the true mock bulk flow, demonstrating that the NN is sensitive to information on ``super-survey'' scales encoded in the RSDs. When applied to 2MRS, the NN successfully recovers the main known clusters, some of which are partially in the Zone of Avoidance. The reconstructed bulk flows in spheres of different radii less than $100\,h^{-1}\,\mathrm{Mpc}$ are in good agreement with a previous 2MRS analysis that required an additional external bulk flow component inferred from directly observed peculiar velocities. The NN-reconstructed peculiar velocity of the Local Group closely matches the observed Cosmic Microwave Background dipole in amplitude and Galactic latitude, and only deviates by $18^\circ$ in longitude. The NN-reconstructed fields are publicly available.}
{}


\maketitle
%

\section{Introduction}
\label{sec:introduction}
The cornerstone of modern cosmology rests on gravity as the driving force behind the formation of the large-scale structure (LSS) observed in the Universe. Initially, minuscule fluctuations, seeded by quantum fluctuations, were magnified through gravitational instability, culminating in the intricate cosmic structures we observe today. The mass density field shaping this cosmic tapestry is predominantly governed by dark matter (DM), which contributes roughly $85\%$ of the mass and thus the gravitational influence, while ordinary baryonic matter constitutes the remaining $15\%$.

Galaxies trace the underlying density field, congregating in elongated filaments that connect clusters surrounded by vast underdense regions (voids), collectively forming what is known as the cosmic web. Understanding the genesis and evolution of the LSS is paramount for advancing cosmological knowledge and deciphering the Universe's composition, as well as any deviations from the standard model.

A crucial research objective lies in reconstructing the density and velocity fields from the observed galaxy distribution. The reconstructed galaxy density field offers insights into the large-scale distribution of DM, contingent upon establishing a ``biasing relation'' that links galaxies to DM. Moreover, within the standard paradigm, a simple relationship exists between peculiar velocity and density, allowing for the inference of the velocity field from the reconstructed density field. This inferred velocity field can then be compared with the observed velocities of galaxies, facilitating the validation of cosmological models and refining our understanding of how galaxies populate DM halos and evolve over cosmic epochs.

More ambitiously, discrepancies between the inferred and observed fields may indicate deviations from standard gravity theory. These deviations could potentially overlap with the biasing relation, making a reliable reconstruction vital for disentangling the consequences of nonstandard gravity from biasing effects. Achieving this requires the development of sophisticated algorithms and techniques capable of accurately mapping density and velocity fields from observational data while rigorously addressing potential biases and uncertainties.

Therefore, substantial effort has been invested in developing reconstruction techniques for inferring cosmological velocity and density fields from observations. The two main types of observational data are (i) catalogs of peculiar velocities of galaxies and (ii) surveys of galaxy redshifts and angular positions on the sky. In this paper we address the latter.

Any reconstruction method of the true mass density from redshift surveys must face the following three main challenges. First, the finite number of observable galaxies introduces shot noise, constituting the primary source of random error. Since only galaxies above a specific flux limit are observed, the dataset is inherently incomplete. To mitigate this issue, each galaxy is weighted by the survey's selection function. This adjustment compensates for the survey's incompleteness at varying distances, enabling a more precise statistical analysis of the galaxy distribution.

Second, galaxies reside within DM halos, and their properties and distribution are influenced by the complex accretion and merging history of these halos. Consequently, the biasing relation between the observed galaxy distribution and the underlying DM density field is complex and nonlinear. Establishing a model for this biasing relation is crucial for any attempt to accurately reconstruct the DM density field from the observed distribution of galaxies.

Third, a galaxy's redshift is affected by its peculiar velocity, in addition to the Hubble expansion velocity. This leads to anisotropies in the galaxy distribution as observed in redshift space rather than real-distance space. These redshift-space distortions (RSDs) introduce systematic deviations from the true galaxy distribution. Correcting for these distortions is crucial to accurately infer the underlying mass density field. On large scales (larger than a few megaparsec), the peculiar velocity field is coherent with the underlying density and hence causes an enhancement of density fluctuations relative to real space. This anisotropic enhancement can actually be used to constrain cosmological parameters. On small scales, random galaxy motions stretch their apparent distribution along the line of sight in redshift space, creating elongated structures known as ``Fingers of God'' \citep{jackson_critique_1972,sargent_statistical_1977,tully_nearby_1978}. Both coherent flows and random motions need to be properly addressed in any successful reconstruction.

Prior to the significant growth and widespread adoption of machine learning in various fields, including cosmology, traditional methods for analyzing LSS observations primarily relied on specific assumptions that could be formulated either analytically or numerically in a compact, well-defined manner. The reconstruction of velocity fields from redshift surveys, for example, required specific density-velocity correlations derived from approximate dynamics. Such correlations were based on models found in linear theory \citep[e.g.][]{peebles_large-scale_1980} or quasi-linear approaches, such as the Zel'dovich approximation \citep{zeldovich_gravitational_1970, nussetal91}, 2LPT \citep{moutarde_precollapse_1991,bouchet_weakly_1992,Buchert1993,Gramann1993,bouchet_perturbative_1995,Zheligovsky2014}, and the least action principle \citep{Peebles_tracing_orbits_1989, nusser_branchini_2000}.
 
To mitigate the effect of shot noise, most reconstruction methods employ a smoothing of the discrete galaxy distribution with an extended (e.g. Gaussian) kernel to obtain a galaxy density field \citep[e.g.][]{yahil_redshift_1991,carrick_cosmological_2015,boruah_cosmic_2020}. Another popular technique is applying a Wiener filter (WF) \citep{wiener_extrapolation_1949}, which yields the linear minimum variance estimator of the true density\,/\,velocity fields given the observed data \citep[e.g.][]{zaroubi_wiener_1995,fisher_wiener_1995,webster_wiener_1997,schmoldt_density_1999,erdogdu_reconstructed_2006,lilow_constrained_2021}. For the special case of Gaussian data and true fields, the WF coincides with the mean and maximum of the posterior distribution of the true fields given the data. A different approach to relate the distributions of noisy, biased tracers and the underlying matter field is based on optimal transport theory \citep{brenier_reconstruction_2003,nikakhtar_optimal_2023,nikakhtar_displacement_2024}. Another widely used class of reconstruction methods are hierarchical Bayesian models \citep[e.g.][]{jasche_physical_2019,kitaura_cosmic_2021}, which draw samples from a posterior distribution that combines different models for gravitational dynamics, RSDs, and biasing relations in a forward-modelling approach. This approach allows for the description of nonlinear relations and non-Gaussian statistics but is usually computationally expensive and still relies on explicit (approximate) modelling assumptions.

The primary advantage of neural networks (NNs) over traditional approaches lies in their ability to infer the relationship between the underlying true fields and the observed LSS data without predefining the functional form of this relationship. Thus, NN algorithms particularly excel at processing training data encompassing a broad spectrum of physical effects. These effects are often too complex for analytical models to accurately capture or computationally too expensive to be evaluated numerically during the inference process. Recent years have therefore seen a rise in studies where NNs are trained to reconstruct density and velocity fields from observations \citep[e.g.][]{wu_cosmic_2021,shallue_reconstructing_2023,ganeshaiah_veena_large-scale_2023,qin_reconstructing_2023,chen_effective_2023,wu_ai-assisted_2023,legin_posterior_2024,chen_constructing_2024}.

In a previous paper, \citet[hereafter \citetalias{ganeshaiah_veena_large-scale_2023}]{ganeshaiah_veena_large-scale_2023}, we have assessed the applicability of an NN for the reconstruction of 3D matter density and velocity fields from a sample of observed galaxies. For that purpose, a simplified set of training data was used, employing 2LPT dynamics, imposing RSDs along the $z$- rather than radial direction, and not including any galaxy biasing or survey selection effects. This proof-of-concept demonstrated that an NN is capable of efficiently learning to evaluate the non-Gaussian mean posterior estimate of the true fields given the observations, yielding a consistently better reconstruction accuracy than a reference WF.

In the current paper, our objective is to apply this NN framework to the Two-Micron All-Sky Redshift Survey (2MRS) \citep{huchra_2mass_2012, macri_2mass_2019} in order to obtain high-quality reconstructions of the 3D matter density and velocity fields in the local Universe. To achieve this, we used the high-resolution N-body simulations of the Quijote suite \citep{villaescusa-navarro_quijote_2020} to extract training data that incorporate accurate gravitational dynamics and mimic the survey characteristics of 2MRS as closely as possible.

The structure of the paper is as follows. In \cref{sec:data}, we provide a brief overview of 2MRS, concentrating on its key properties relevant to the generation of mock data. Additionally, this section introduces the Quijote simulations and describes their use in creating mock data that emulate the characteristics of the actual 2MRS. The generated mock data are intended for the training of our NN, detailed in \cref{sec:neural_net}. Comprehensive testing of the NN, utilizing a validation set of mock data, is documented in \cref{sec:validation}. The outcomes of applying the NN to the actual 2MRS data are presented and examined in \cref{sec_2mrs}. The paper concludes with \cref{sec:conclusion}, offering a summary and discussion of the findings.

\section{Data}
\label{sec:data}
\subsection{2MRS}
\label{sec:data:2MRS}
The 2MRS is a flux-limited survey with a $K_s$-band magnitude of $K_s \leq 11.75$, providing sky positions and spectroscopic redshifts for $44,572$ galaxies \citep{huchra_2mass_2012,macri_2mass_2019}. Its footprint covers $91\%$ of the sky, only missing a thin band covered by the Milky Way, the so-called Zone of Avoidance (ZoA).

In our reconstruction, we consider a spherical volume with a radius of $r_{\mathrm{max}} = 200\,h^{-1}\,\mathrm{Mpc}$, encompassing $98\%$ of the 2MRS galaxies. Outside of the ZoA, the observable fraction of galaxies follows a selection function $\phi(r)$ that solely depends on the distance from the observer. To robustly estimate $\phi(r)$, we adopt the methodology outlined in \citet{lilow_constrained_2021}. This involves imposing a partial volume limit at a distance of $30 \,h^{-1}\,\mathrm{Mpc}$ and utilizing an $F/T$-estimator \citep{davis_survey_1982,branchini_linear_2012}. The number of galaxies within $r_{\mathrm{max}}$ satisfying the partial volume limit is $N^\mathrm{g} = 42648$. However, unlike \citet{lilow_constrained_2021}, we directly employ individual observed galaxy redshifts in the Cosmic Microwave Background (CMB) frame rather than redshifts associated with galaxy groups. While grouping galaxies is typically undertaken to mitigate Fingers-of-God effects in redshift space, we found it unnecessary for generating high-quality NN reconstructions.

The selection-function-corrected mean galaxy number density within the reconstruction volume is given by
\begin{equation}
    \bar{n}^\mathrm{g} = \frac{3}{4 \uppi \, r_\mathrm{max}^3 \, F_\mathrm{sky}} \, \sum_{a=1}^{N^\mathrm{g}} \, \frac{1}{\phi(s_a)} \approx 1.83 \times 10^{-2} \, h^3 \, \mathrm{Mpc}^{-3} \,,
    \label{eq:2MRS_galaxy_number_density}
\end{equation}
where $F_\mathrm{sky} \approx 0.91$ is the fraction of the sky not obscured by the ZoA, and $\phi(s_a)$ is evaluated at the comoving redshift-space distances $s_a$ of the observed galaxies.

The 2MRS galaxy number count field used as input to the NN reconstruction was obtained by defining a regular $128\times128\times128$ cubic grid of $400 \,h^{-1}\,\mathrm{Mpc}$ side length enclosing the reconstruction volume and assigning each of the $N^\mathrm{g}$ galaxies to the grid point nearest to the galaxy's redshift-space coordinate $\vec{s}_a$.

\subsection{Mock data}
\label{sec:data:mock}
The accuracy of an NN reconstruction hinges upon the quality of the training data it receives. To generate mocks that employ an accurate model of nonlinear DM dynamics on scales $\gtrsim 3\,h^{-1}\,\mathrm{Mpc}$, we used the Quijote N-body simulations suite, which aims at providing a sufficiently large number of high-quality simulations for training machine learning applications \citep{villaescusa-navarro_quijote_2020}. For this paper, we used the 100 high-resolution fiducial cosmology runs, each containing $1024^3$ equal-mass particles in a cubic box with side length $1000 \,h^{-1}\,\mathrm{Mpc}$. The cosmological parameters used are $\Omega_\mathrm{m} = 0.3175$, $\Omega_b = 0.049$, $h = 0.6711$, $n_s = 0.9624$, $\sigma_8 = 0.834$, which are in good agreement with the \textit{Planck} 2018 constraints \citep{aghanim_planck_2020-1}.

From every simulation box we extracted $64$ partially overlapping cubic sub-boxes of side length $400\,h^{-1}\,\mathrm{Mpc}$, resulting in a total number of $6400$ mocks. Among these, $5760$ were designated for training data, while the remaining $640$ were set aside for validation purposes. Although neighboring mocks are not completely independent, as they overlap by $150 \,h^{-1}\,\mathrm{Mpc}$, we found this to result in a better NN training convergence than using fewer nonoverlapping mocks. In part, this is due to the independently created mock observations, as described in the following.

To ensure that the mock observations are as close as possible to the actual 2MRS characteristics, we accounted for RSDs, ZoA, selection function $\phi(r)$ and galaxy bias. Due to the flux-limit of 2MRS, the minimum luminosity of observable galaxies increases with distance, leading to a radially growing galaxy bias. To reproduce this in our mocks, we assumed that a galaxy can only form in regions where the matter density $1 + \delta$ is above a radially increasing threshold $\rho^\mathrm{th}(r)$. The expected number of formed galaxies is then proportional to the modified density
\begin{equation}
    1 + \delta^\mathrm{th}(\vec{r}) =
    \begin{cases}
        \bigl(1 + \delta(\vec{r})\bigr) \, \mathcal{N}(r) &\text{if $1 + \delta(\vec{r}) \geq \rho^\mathrm{th}(r)$,} \\
        0 &\text{else,}
    \end{cases}
    \label{eq:threshold_density}
\end{equation}
where the normalization factor $\mathcal{N}(r)$ was chosen such that the ensemble average of $1 + \delta^\mathrm{th}(\vec{r})$ at any position $\vec{r}$ is unity. The threshold values $\rho^\mathrm{th}(r)$ for different radii were calibrated such that the resulting galaxy density fluctuation amplitude $\sigma_8^\mathrm{g}(r)$ in our mocks matches that of 2MRS (computed using the method described in Appendix A of \citet{lilow_constrained_2021}).

The mock matter density and peculiar velocity fields, used as NN targets, and the mock galaxy number count fields, used as NN inputs, were then obtained as follows:
(i) In each sub-box, the mean simulation particle density as well as the mean and variance of the particle velocities were computed on a regular $128\times128\times128$ grid, using a Cloud-in-Cell (CIC) assignment scheme. The mean particle density and velocity grids constituted the mock matter density and peculiar velocity fields, respectively.
(ii) For each grid cell $j$ at a distance $r_j \leq r_\mathrm{max}$ to the sub-box center, a random number $N_j$ of observed galaxies was drawn from a Poisson distribution $\mathrm{Pois}(N_j\,|\,\bar{N}_j)$ around the expected number of observable galaxies in that cell, $\bar{N}_j = (1 + \delta^\mathrm{th}_j) \, \phi(r_j) \, \bar{n}^\mathrm{g} \, V$. Here, $1 + \delta^\mathrm{th}_j$ is the grid cell's modified matter density according to \cref{eq:threshold_density}, $\bar{n}^\mathrm{g}$ is the selection-function-corrected mean 2MRS galaxy number density \cref{eq:2MRS_galaxy_number_density} and $V$ is the volume of a grid cell.
(iii) The real-space Galactic coordinates of the $N_j$ galaxies were each set to an independent uniformly random position $\vec{r}_a$ within the grid cell $j$. If $\vec{r}_a = 0$ or if the galaxy's Galactic longitude and latitude lie within the ZoA, this galaxy was discarded.
(iv) Each remaining galaxy was assigned a peculiar velocity $\vec{v}_a$ drawn from a Gaussian distribution whose mean and variance are the CIC-assigned mean and variance of the simulation particle velocities, linearly interpolated between grid points at the galaxy position $\vec{r}_a$.
(v) The observed galaxy redshift was given by $1 + z^\mathrm{obs}_a = (1 + z_a) \, (1 + \frac{\vec{v}_a \cdot \vec{r}_a}{c \, r_a})$, where $z_a$ is the cosmological redshift corresponding to the galaxy's comoving distance, $r_a = D_\mathrm{com}(z_a)$, and $c$ is the speed of light. If $z^\mathrm{obs}_a < 0$, this galaxy was discarded. Otherwise, the galaxy's redshift-space coordinate was $\vec{s}_a = D_\mathrm{com}(z^\mathrm{obs}_a) \, \frac{\vec{r}_a}{r_a}$.
(vi) Finally, the mock galaxy number count field $N^\mathrm{obs}$ was obtained by assigning each remaining galaxy to the grid point nearest to the galaxy's redshift-space coordinate $\vec{s}_a$.

\section{Neural network}
\label{sec:neural_net}
An NN is a computational model with adjustable weights and biases, fine-tuned by minimizing a loss function, to predict targets from input data. It consists of layers, each with multiple neurons, where weights adjust connection strengths and biases set activation thresholds.

Our main goal is to train an NN to infer the 3D matter density and peculiar velocity fields from noisy observed galaxy number counts, described in detail in \cref{sec:data}. For this purpose, we trained two separate NNs: one for the density field and another for the velocity field.

We utilized an autoencoder with U-Net architecture, which has proven effective for this reconstruction task in \citetalias{ganeshaiah_veena_large-scale_2023}. It comprises an encoder, which compresses the input to a lower-dimensional latent space representation, and a decoder, which reconstructs the target from the latent space representation. Both input and target fields were represented as $128 \times 128 \times 128$ grids, and the latent space dimension was $(8 \times 8 \times 8)_{\mathrm{grids}} \times 128_{\mathrm{filters}}$. In summary, the encoder used a series of 3D convolution steps with ReLU activation and max pooling to reduce the grid size; the decoder employed a symmetric series of 3D transpose convolution steps with ReLU activation to expand the grid size again. Skip connections between the encoder and decoder helped maintain the spatial feature localization and improve the training convergence. The density NN included a final ReLU activation, ensuring that no negative densities are generated, whereas the velocity NN used a linear final activation. The precise NN architecture, including the numbers of convolutional filters, is detailed in \citetalias{ganeshaiah_veena_large-scale_2023}. The only modification made in the present work was to reduce the depth of the NN by one step (in \citetalias{ganeshaiah_veena_large-scale_2023} the latent dimension was $(4 \times 4 \times 4)_{\mathrm{grids}} \times 256_{\mathrm{filters}}$), resulting in a total number of $\sim 1.6\times 10^6$ trainable parameters. This change was found to reduce overfitting without sacrificing reconstruction accuracy.

Before being fed into the NN, the input data underwent preprocessing to ensure that the values are predominantly constrained within the range of 0 to 1. The input galaxy number counts and the target density fields were both divided by a factor of 40, while the target velocity fields were divided by 300, approximately matching the standard deviation per velocity component.

\begin{figure*}
    \centering
    \includegraphics[width=\textwidth]{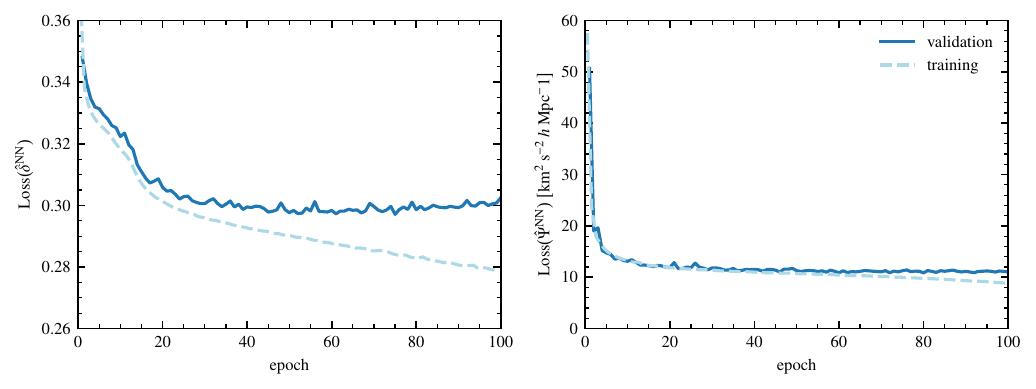}
    \caption{Loss functions as a function of training epochs. \textit{Left panel:} loss function of the reconstructed density field, given in \cref{eq:loss_density}, in the validation (solid blue) and training (dashed light blue) sets of mocks. \textit{Right panel:} same for the peculiar velocity potential field, given in \cref{eq:loss_velocity}.}
    \label{fig:loss_function}
\end{figure*}

As discussed in \citetalias{ganeshaiah_veena_large-scale_2023}, the choice of loss function determines the type of statistical estimator learned by the NN. Specifically, we are interested in computing the mean posterior estimate $\hat{T}_j \in \{\hat{\delta}_j^\mathrm{NN}, \hat{\vec{v}}_j^\mathrm{NN}\}$ of the target fields $T_j \in \{\delta_j^\mathrm{true}, \vec{v}_j^\mathrm{true}\}$ given an input field $I_j = N_j^\mathrm{obs}$, where $j = 1,\dotsc,M_\mathrm{grid}$ labels the individual grid cells. This is achieved by employing a (weighted) mean squared error loss function,
\begin{equation}
    \mathrm{Loss}(\hat{T}) = \underbrace{\sum_{I,T} \, P(I,T)}_{\sum_I P(I) \, \sum_T P(T\,|\,I)} \! \sum_{j=1}^{M_\mathrm{grid}} \, \omega_j \, \Bigl(T_j - \hat{T}_j(I)\Bigr)^2 \,,
\end{equation}
where $\omega_j$ is a grid-cell-dependent weighting factor and $P(I,T)$ is the joint probability distribution of input and target fields. Via Bayes' theorem, the latter can be split into the evidence $P(I)$ and the posterior $P(T\,|\,I)$. Minimizing this loss function yields
\begin{equation}
    0 = \frac{\updelta \mathrm{Loss}(\hat{T})}{\updelta \hat{T}_j(I)} = -2 \, \omega_j \, P(I) \, \sum_T \, P(T\,|\,I) \, \Bigl(T_j - \hat{T}_j(I)\Bigr)
    \label{eq:general_weighted_mse_loss}
\end{equation}
and thus the desired mean posterior estimate
\begin{equation}
    \hat{T}_j(I) = \sum_T \, P(T\,|\,I) \, T_j = \langle T_j \,|\, I\rangle \,.
    \label{eq:mean_posterior_estimate_definition}
\end{equation}
Formally, any $\hat{T}$-independent choice of the weighting factor $\omega_j$ results in the same estimator.\footnote{It may even depend on the target and input fields, $T$ and $I$.} But since during training the loss function is minimized numerically via stochastic gradient descent \citep{goodfellow_deep_2016}, the choice of weighting can influence the numerical convergence and thus the actually learned estimator. Also, we note that in practice the infinite sum over all possible samples from $P(I,T)$ in \cref{eq:general_weighted_mse_loss} needs to be approximated by a sum over a finite set of training samples.

For the training of the density reconstruction NN, we found that a weighting with the survey selection function, $\omega_j \propto \phi(r_j)$, improves the training convergence. This choice was motivated by the fact that $\phi$ is proportional to the inverse variance of the observed galaxy shot noise. We hence minimized the loss function
\begin{equation}
    \mathrm{Loss}\bigl(\hat{\delta}^\mathrm{NN}\bigr) = \frac{1}{M_\mathrm{train} M_\mathrm{grid}} \, \sum_{\alpha=1}^{M_\mathrm{train}} \sum_{j=1}^{M_\mathrm{grid}} \, \phi(r_j) \, \Bigl(\delta_j^{\mathrm{true},\alpha} - \hat{\delta}_j^{\mathrm{NN},\alpha}\Bigr)^2 \,,
    \label{eq:loss_density}
\end{equation}
where $\alpha = 1,\dotsc,M_\mathrm{train}$ labels the individual mock realizations used for training.

For the training of the velocity reconstruction NN, one could use an equivalent loss function for the individual velocity components, but that would require an NN with three output channels or three separate NNs, since the ZoA (which has different widths for different Galactic latitudes) breaks the symmetry between the velocity components. However, this can potentially result in the prediction of a spurious rotational component in the reconstructed velocity field, which we know to be absent on the scales $\gtrsim 3 \,h^{-1}\,\mathrm{Mpc}$ probed in our setup.\footnote{In principle, an NN predicting the individual velocity components should learn their irrotationality during training, since by construction the mean posterior estimate \cref{eq:mean_posterior_estimate_definition} of irrotational true fields is itself irrotational. In practice, however, the trained NN is not guaranteed to precisely preserve this symmetry, as the NN only approximates the mean posterior estimate.}

Therefore, we instead imposed the reconstruction of an irrotational velocity field by assuming a potential flow, $\hat{\vec{v}}^\mathrm{NN} = \vec{\nabla} \hat{\Psi}^\mathrm{NN}$, and training the NN to reconstruct the velocity potential $\hat{\Psi}^\mathrm{NN}$. This also improved the training convergence compared to learning to reconstruct the individual velocity components directly. Beyond that, the convergence was empirically found to be further improved by using a weighting $\omega_j \propto \phi(r_j) / r_j$, which gives even more weight to grid cells at smaller distances than the weighting in the density loss function. The velocity (potential) loss function thus reads
\begin{equation}
    \mathrm{Loss}\bigl(\hat{\Psi}^{\mathrm{NN}}\bigr) = \frac{1}{M_\mathrm{train} M_\mathrm{grid}} \, \sum_{\alpha=1}^{M_\mathrm{train}} \sum_{j=1}^{M_\mathrm{grid}} \, \frac{\phi(r_j)}{r_j} \, \Bigl(\vec{v}_j^{\mathrm{true},\alpha} - \vec{\nabla} \hat{\Psi}_j^{\mathrm{NN},\alpha}\Bigr)^2 \,,
    \label{eq:loss_velocity}
\end{equation}
where the potential gradient is computed via symmetric finite differences.

The NNs were implemented in Keras \citep{keraschollet2015} with a TensorFlow backend \citep{abadi_tensorflow_2016}. Using the AMSGrad variant of the Adam parameter optimizer \citep{kingma_adam_2017,reddi_convergence_2019} with learning rate $10^{-4}$ and a batch size of 4 mock realizations per gradient update, training the density and velocity (potential) reconstruction NNs for 100 epochs took around $12$ hours each on an NVIDIA 3090 GPU. Once trained, the NN reconstruction of either target field for a given galaxy number count field takes just a fraction of a second.

\begin{figure*}
    \centering
    \includegraphics[width=0.93\textwidth]{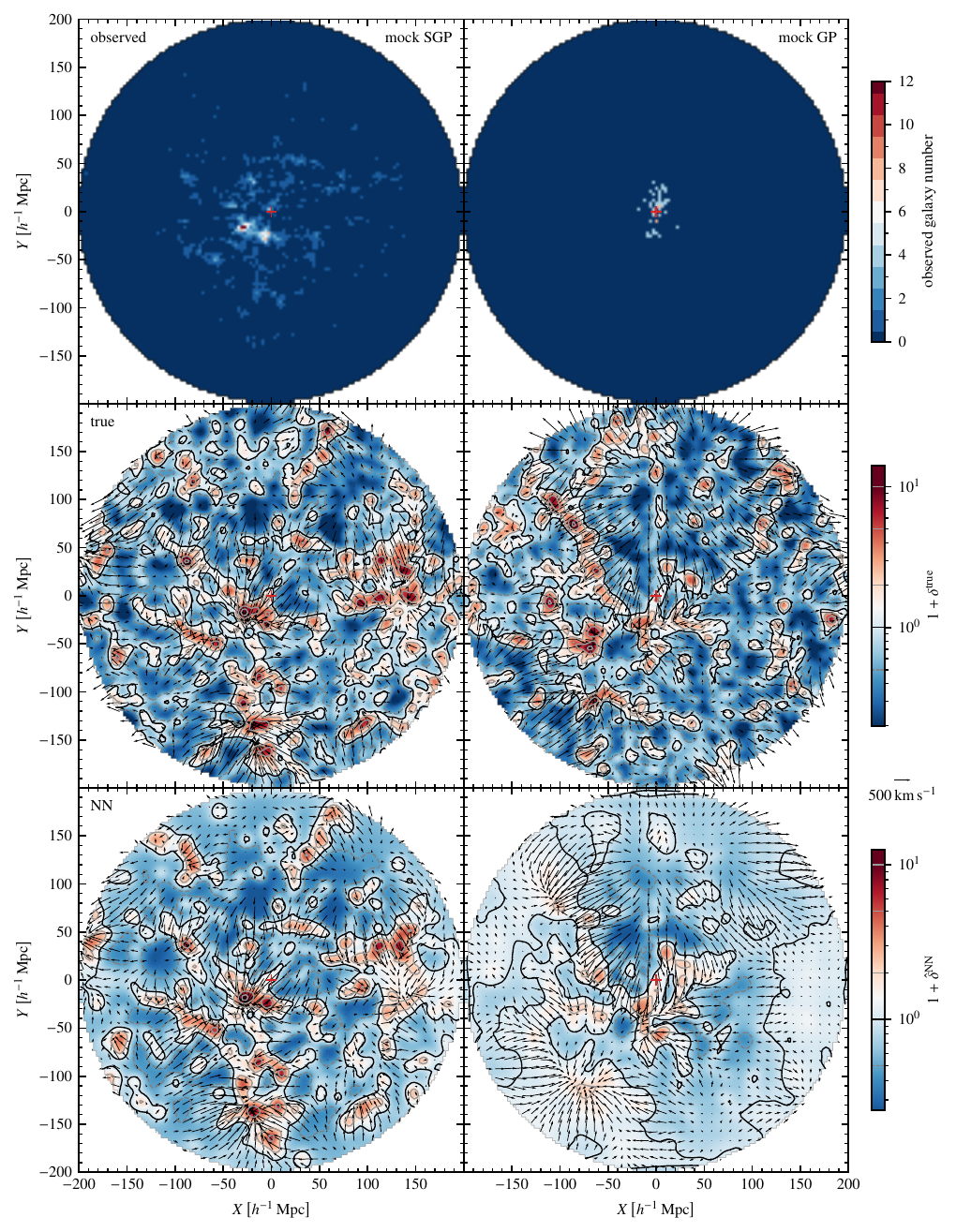}
    \caption{Observed, true and reconstructed fields in a slice through the Supergalactic (left) and Galactic (right) planes for one of the validation mocks. \textit{Top row:} observed galaxy numbers. The red cross marks the position of the observer at the origin. \textit{Middle row:} true density (heat map and contours) and peculiar velocity (arrows) fields. The values represented by the contours are marked in the color bar. A reference arrow representing a velocity of $500 \,\mathrm{km}\,\mathrm{s}^{-1}$ is shown on the right. \textit{Bottom row:} corresponding NN-reconstructed fields.}
\label{fig:mock_field_slices}
\end{figure*}

The loss functions as a function of the training epoch, evaluated for both the training and the validation set are shown in \cref{fig:loss_function}. For the density (left panel), the training loss drops quickly at first before settling into a slow and steady decrease after $\sim 20$ epochs. The validation loss initially follows the training loss, then plateaus after $\sim 40$ epochs, fluctuates around its minimum for some time and eventually starts to weakly increase again after $\sim 80$ epochs. For the velocity (right panel), a qualitatively similar behavior is seen, but with a faster initial drop in training loss, followed by a much slower steady decrease after $\sim 10$ epochs. The velocity validation loss stabilizes after $\sim 40$ epochs, showing only minor fluctuations but no sign of increasing again within the 100 epochs. For the reconstructions in the subsequent sections we employ the NN models at the epochs with the minimal validation loss, epoch 75 for the density and epoch 60 for the velocity.

\section{Validating the reconstruction using the mocks}
\label{sec:validation}
In the following, we assess the reconstruction quality achieved by the trained NN by performing a number of tests on the validation set of 640 mocks, described in \cref{sec:data:mock}. For some of those tests, we compare the NN results against the widely used and robust WF, serving as a benchmark. We specifically employ the WF implementation presented in \citet{lilow_constrained_2021}, which is designed for all-sky surveys such as 2MRS. It combines the WF with a linear RSD correction, and exploits the isotropy of the survey by expanding the WF in spherical Bessel functions and spherical harmonics \citep{fisher_wiener_1995}. This requires filling up the ZoA by copying galaxies from adjacent regions \citep{yahil_redshift_1991}, which only has a minor impact on the WF reconstruction results outside of the ZoA, though, as the ZoA only blocks a small portion of the sky.

For all tests, the reconstructed as well as true density and velocity fields were first smoothed using a Gaussian window of width $r_{\mathrm{smooth}} = 3 \,h^{-1}\,\mathrm{Mpc}$, approximately matching the spatial resolution of the employed field grids, to reduce aliasing effects.

\subsection{Visual inspection}
\label{sec:validation:visual_inspection}
In \cref{fig:mock_field_slices}, we present the observed, true and NN-reconstructed fields for a single example mock, chosen for its qualitative similarity to the distribution of structures seen in 2MRS. We are showcasing the fields both in a slice through the mock Supergalactic plane (SGP, left panels) and the mock Galactic plane (GP, right panels). The field values on those slices were obtained by linear interpolation between the grid values, resulting in an effective slice thickness of $\sim 3 \,h^{-1}\,\mathrm{Mpc}$, approximately matching the grid resolution.

For the density in the SGP, the visual inspection reveals a good agreement between the true and reconstructed fields. Large-scale structures, particularly those near the center where galaxies sample the matter distribution densely, are robustly reconstructed. Dense regions are faithfully reproduced, and the larger filamentary structures within the slice are also captured in the reconstruction. Even in the more sparsely sampled outer regions, the NN effectively reconstructs structures to a significant extent.

However, we note a slight discrepancy in the representation of underdense regions, such as voids. They appear slightly less underdense in the reconstructed fields compared to the true fields. This discrepancy is attributed to the NN-reconstructed fields representing the mean of all possible true realizations (mean posterior estimate) and has already been seen and discussed in \citetalias{ganeshaiah_veena_large-scale_2023}.

In the GP, which cuts directly through the ZoA, the top-right panel only reveals a few galaxies near the center, where the $\sim 3 \,h^{-1}\,\mathrm{Mpc}$ thick slice extends beyond the ZoA (which spans a fixed range in latitudes rather than a fixed distance perpendicular to the GP). Despite the diminished number of tracers, the reconstruction in the GP (bottom-right panel) effectively captures the most dominant structures on large scales, even at considerable distances from the observer. Particularly, the large filamentary structure seen in the top-left quadrant of the true field (middle-right panel) is well-captured in the reconstructed field (the bottom-right panel). Naturally, though, the reconstruction is generally more successful at distances closer to the observer where the width of the ZoA is smaller.

\begin{figure*}
    \centering
    \includegraphics[width=\textwidth]{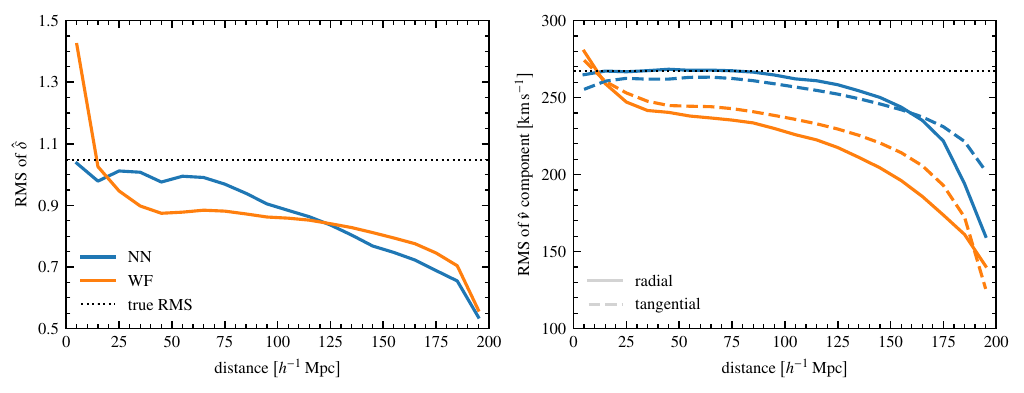}
    \includegraphics[width=\textwidth]{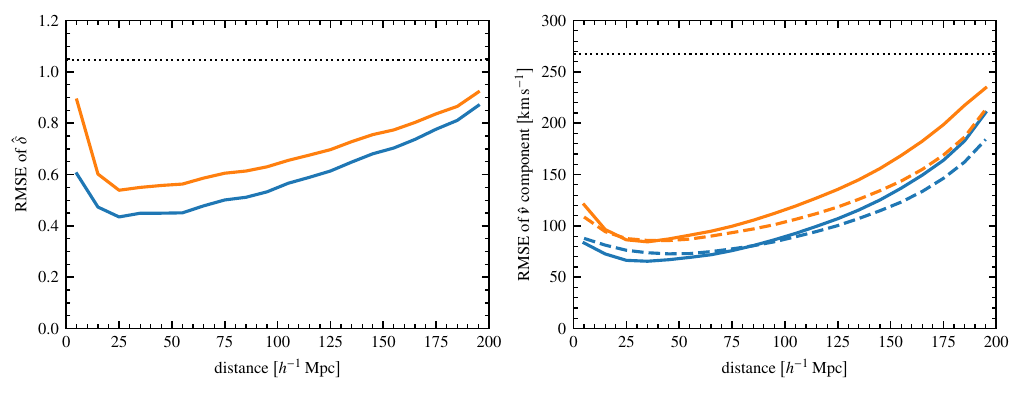}
    \caption{Root mean square (top) and root mean squared error (bottom) of the reconstructed fields as a function of distance. \textit{Top-left panel:} root mean square of the reconstructed density contrast for the NN (blue) and WF (orange) in the validation mocks. The RMS is averaged over spherical shells of width $10 \,h^{-1}\,\mathrm{Mpc}$. For reference, the RMS of the true field is marked by the horizontal dotted line. \textit{Top-right panel:} same for the radial (solid) and tangential (dashed) peculiar velocity components. \textit{Bottom row:} same for the root mean squared error.}
    \label{fig:mock_rms_rmse}
\end{figure*}

The velocity fields are overlaid in \cref{fig:mock_field_slices} as arrows. The coherence between density and velocity is apparent in all relevant panels, as expected by the gravitational instability theory for structure formation. It effectively captures the overall convergence in regions of higher density and divergence in regions of lower density. The velocity flow seen in the reconstructed fields is remarkably close to the flows in the true fields, especially in the SGP. But even in the sparsely sampled GP, the true and reconstructed velocity fields show a good agreement in all but a few, typically outer regions. This higher reconstruction accuracy of the velocity field compared to the density field in the GP is expected to be a consequence of the significantly larger correlation length of the density field, which allows the ZoA to be partially bridged.

\subsection{RMS and RMSE versus distance}
\label{sec:validation:rms_rmse}
For a more quantitative assessment of the reconstructions, we computed the root mean square (RMS) of the reconstructed fields and the root mean squared error (RMSE) between the reconstructed and true fields. In \cref{fig:mock_rms_rmse} we show these quantities, averaged across all validation mocks, for both the density and velocity fields as a function of the distance from the observer at $r=0$. Both the RMS and RMSE values were computed in spherical shells, each spanning $10 \,h^{-1}\,\mathrm{Mpc}$ in width and extending out to a distance of $200 \,h^{-1}\,\mathrm{Mpc}$.

For $r\lesssim 100\,h^{-1}\,\mathrm{Mpc}$, the RMS of the NN-reconstructed density (blue curve in top-left panel) is impressively close to the RMS value obtained from the true density fields (dotted horizontal line). At larger radii, the RMS becomes significantly lower than the true value, due to the reduced number density of galaxies at these distances. Since the NN-reconstructed density approximates the mean of all true fields consistent with the observations, its RMS is expected to lie below the true value.

Although the performance of the WF (orange curve) matches that of the NN at large distances, it is clearly less successful at consistently approximating the true value for smaller distances. In fact, it even overshoots the true value close to the center by about $30\%$.

A similar behavior is observed for the RMS of the radial and tangential velocity components (top-right panel). At $r\lesssim 100\,h^{-1}\,\mathrm{Mpc}$, the NN successfully recovers the RMS for both components with high accuracy. The radial component (solid blue curve) nearly converges exactly to the RMS of the true velocity field components (dotted line), while the tangential component (dashed blue curve), falls only slightly short of reaching the true RMS.

The bottom panels in \cref{fig:mock_rms_rmse} plot curves of the RMSE versus distance. Here, lower RMSE values indicate a closer match between the reconstructed fields and the true ones.

\begin{figure*}
    \centering
    \includegraphics[width=\textwidth]{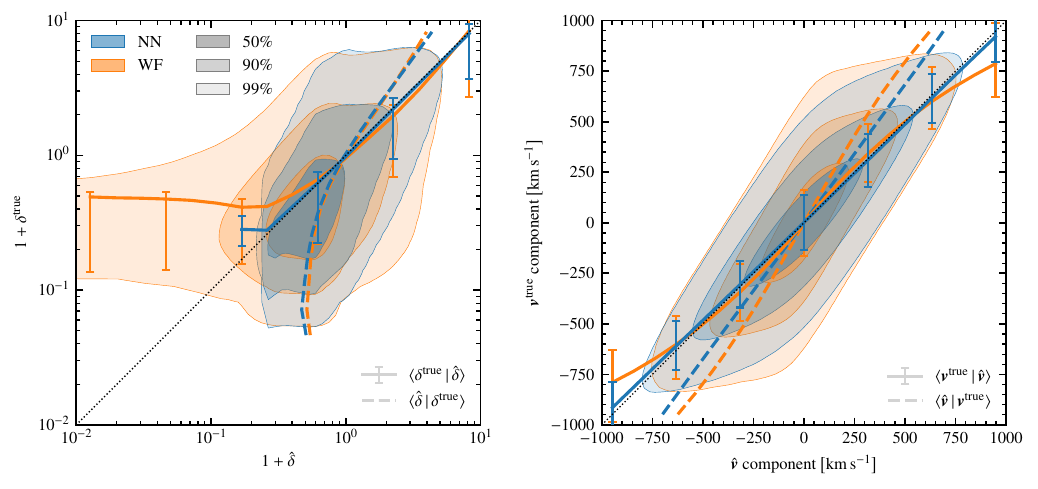}
    \caption{Distribution of true versus reconstructed field values. \textit{Left panel:} distribution of true versus reconstructed density values for the NN (blue) and WF (orange) reconstructions in the validation mocks, represented as shaded areas bounded by the iso-density contours of the distribution. Going from a darker to lighter hue, the shaded areas contain $50\%$, $90\%$, and $99\%$ of all the points, respectively. The conditional mean true values and the $68\%$ most likely true values for a given reconstructed value are shown as solid lines and error bars, respectively. The average reconstructed values for a given true value are shown as dashed lines (understood as a function of the true values). For reference, the diagonal is marked with a dotted line. \textit{Right panel:} same for the peculiar velocity components.}
    \label{fig:mock_true_vs_reconstructed}
\end{figure*}

For the density (bottom-left panel), the RMSE is lowest at $r\approx 25\,h^{-1}\,\mathrm{Mpc}$ and then gradually increases toward the boundary of the reconstruction volume at $r=200\,h^{-1}\,\mathrm{Mpc}$. Overall, the NN estimates consistently exhibits a lower RMSE than the corresponding WF estimates.

At $r \lesssim 25 \,h^{-1}\,\mathrm{Mpc} $, the RMSE increases as we approach the origin. This increase arises primarily due to RSD effects, since the peculiar velocity at the origin induces an artificial dipole enhancement of the density in redshift space. We confirmed that the discrepancy does not exist when the NN is trained on real-space mock observations or when the reconstructed field is smoothed over larger smoothing scales.

The velocity estimates (bottom-right panel) also exhibit a similar trend: low RMSE values at small distances that gradually increase toward the boundary, both for the radial and tangential components. The NN estimates consistently exhibit closer agreement with the true values compared to the WF estimates.

\subsection{Distribution of true versus reconstructed field values}
\label{sec:validation:true_vs_reconstructed}
We highlight the differences between the true and reconstructed field values through a detailed point-by-point comparison in \cref{fig:mock_true_vs_reconstructed}, presented as iso-density contours of the distribution of true versus reconstructed values. The contours encompass different fractions of the grid points within the reconstruction volume, as indicated in the figure, revealing detailed characteristics of the distribution.

For the density (left panel), we clearly see that the NN contours (blue) are generally tighter than the WF contours (orange), thus describing a smaller scatter between true and reconstructed densities. Most notably, the contours for the WF extend toward far smaller reconstructed densities than found for the true density, $1+\delta^\mathrm{true} \gtrsim 0.05$. In fact, although not visible in this log-plot, the WF-reconstructed densities, $1+\hat{\delta}^\mathrm{WF}$, sometimes even reach negative values, as the linear WF estimator does not guarantee the positivity of the density. In contrast, the NN-reconstructed densities are constrained to $1+\hat{\delta}^\mathrm{NN} \gtrsim 0.2$. This matches the observation from the visual field inspection in \cref{sec:validation:visual_inspection} that the NN slightly overpredicts the density in very underdense regions, explained by the properties of the mean posterior estimate.

\begin{figure*}
    \centering
    \includegraphics[width=\textwidth]{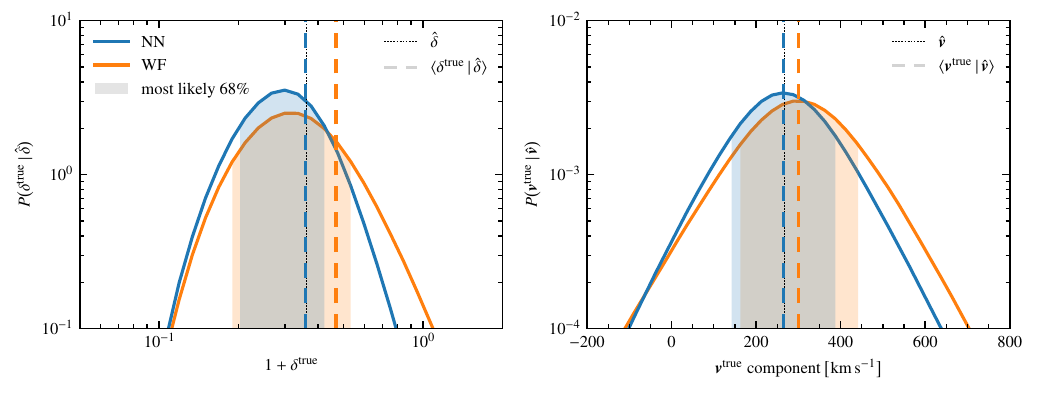}
    \includegraphics[width=\textwidth]{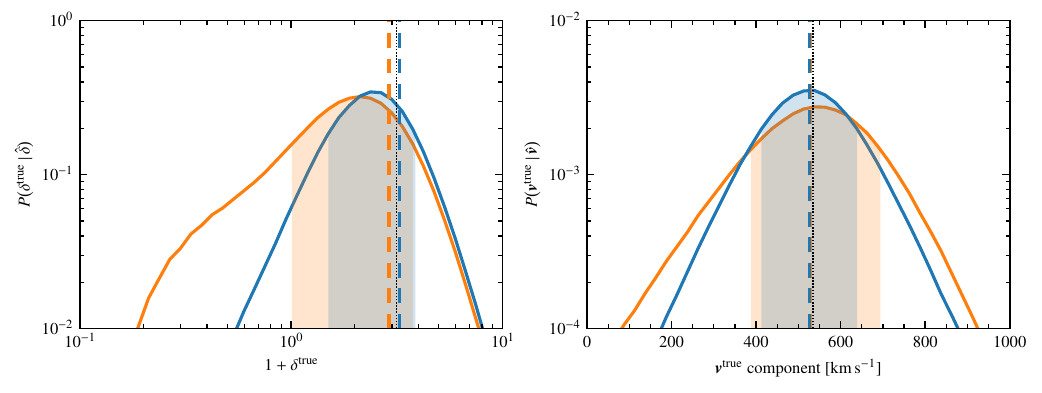}
    \caption{Conditional distributions of true given reconstructed field values. \textit{Top-left panel:} conditional distribution of true densities for a given reconstructed density $1+\hat{\delta}=0.36$ (marked by the vertical dotted line) for NN (blue) and WF (orange) in the validation mocks. The conditional mean true values and the $68\%$ most likely true values are shown as vertical dashed lines and shaded ares, respectively. \textit{Top-right panel:} same for the peculiar velocity components for a given reconstructed velocity component of $\hat{\vec{v}}=267\,\mathrm{km}\,\mathrm{s}^{-1}$. \textit{Bottom row:} same for $1+\hat{\delta}=3.16$ and a component of $\hat{\vec{v}}=534\,\mathrm{km}\,\mathrm{s}^{-1}$.}
    \label{fig:mock_pdf}
\end{figure*}

If the NN-reconstructed density contrast $\hat{\delta}^\mathrm{NN}$ approaches the mean posterior, in accordance with \cref{eq:mean_posterior_estimate_definition}, then it is expected that the conditional mean of $\delta^\mathrm{true}$ given $\hat{\delta}^\mathrm{NN}$ will converge to 
\begin{equation}
    \langle \delta^\mathrm{true} \,|\, \hat{\delta}^\mathrm{NN} \rangle = \hat{\delta}^\mathrm{NN} \,,
    \label{eq:meanNN}
\end{equation}
and the analogous behavior is expected for the velocity. The solid blue line shows the conditional mean computed from the actual fields by averaging the values of $\delta^\mathrm{true}$ in different bins of $\hat{\delta}^\mathrm{NN}$. The result matches the diagonal dotted one-to-one line with almost perfect agreement. The only notable deviation is the smallest bin around $1+\hat{\delta}^\mathrm{NN} = 0.2$. However, it can be seen from the contours that this bin encompasses only a tiny fraction of less than $1\%$ of all points. By comparison, the conditional average of true given WF-reconstructed densities (solid orange line) deviates significantly from the diagonal line for any underdense values and also shows notable discrepancies for overdensities. This is expected because the WF should only reproduce the mean posterior estimate accurately if both the true and observed fields were Gaussian.

The error bars attached to the solid lines mark the range of the most likely $68\%$ true densities given a reconstructed density value. These error bars are asymmetrical around the conditional true mean values, reflecting the asymmetry of the underlying conditional distribution, as discussed in more detail in \cref{sec:validation:pdf}. We see that the error bars for the NN are consistently narrower than those for the WF, especially at the low and high density tails, demonstrating the superior predictive power of the NN.

It is important to note that while \cref{eq:meanNN} holds at least approximately, this is not true for the analogous relation for the reversed conditional mean, $\langle \hat{\delta}^\mathrm{NN} \,|\, \delta^\mathrm{true} \rangle \neq \delta^\mathrm{true}$. That is, the mean posterior estimate is biased in the sense that the conditional average of reconstructed values for a given true value does generally not match that true value. This is demonstrated by the dashed blue line (to be understood as a function of $\delta^\mathrm{true}$), which strongly deviates from the diagonal line. The same is true for the WF (dashed orange line).

We find a similar trend for the peculiar velocity reconstructions in the right panel of \cref{fig:mock_true_vs_reconstructed}. The contours are tighter for the NN compared to the WF fields. The conditional mean of true given reconstructed velocities (solid lines) closely follows the one-to-one line for the NN, but notably deviates from it for the WF, in particular for reconstructed velocity components beyond $\pm 700 \,\mathrm{km}\,\mathrm{s}^{-1}$ as well as close to the true velocity component RMS of approximately $\pm 270 \,\mathrm{km}\,\mathrm{s}^{-1}$. Furthermore, the range of the $68\%$ most likely true velocity values given a reconstructed value is consistently tighter for the NN than for the WF. Lastly, like for the density, the conditional mean of reconstructed given true velocities (dashed lines) significantly deviates from the one-to-one line.

\subsection{Conditional PDF}
\label{sec:validation:pdf}
The error bars in \cref{fig:mock_true_vs_reconstructed} reveal a substantial scatter in the true density and velocity fields when conditioned on their reconstructed values. To better illustrate the statistical nature of this scatter, we now explore the conditional probability distribution function (CPDF) of the true fields for specific given values of the reconstructed density and velocity fields in \cref{fig:mock_pdf}.

For the density, we calculated the CPDFs $P(\delta^\mathrm{true}\,|\,\hat \delta)$ at $1+\hat{\delta}= 0.36$ (top-left panel) and $3.16$ (bottom-left panel). These two values correspond to $-1\,\sigma$ and $2\,\sigma$ fluctuations of the distribution of $\ln(1+{\delta}^\mathrm{true})$. For the velocity component CPDFs $P(\vec{v}^\mathrm{true}\,|\,\hat{\vec{v}})$, we chose $\hat{\vec{v}}$ components of $267\,\mathrm{km}\,\mathrm{s}^{-1}$ (top-right panel) and $534\,\mathrm{km}\,\mathrm{s}^{-1}$ (bottom-right panel), which represent $1\,\sigma$ and $2\,\sigma$ fluctuations of the $\vec{v}^\mathrm{true}$ component distribution. 

\Cref{fig:mock_pdf} clearly reveals that the density CPDFs are skewed in the sense that the conditional mean values $\langle\delta^\mathrm{true}\,|\,\hat{\delta}\rangle$ (dashed vertical lines) are shifted to the right from the corresponding CPDF maxima (the apparent visual symmetry of the CPDFs for the NN is due to the logarithmic scale). This also explains the asymmetry of the $68\%$ most likely true values (shaded areas) relative to the conditional mean, previously seen in the error bars in \cref{fig:mock_true_vs_reconstructed}. For the reconstructed overdensity (bottom-left panel), we furthermore find a notably extended tail of the CPDF for the WF toward low true densities, which is absent for the NN. An NN-reconstructed overdensity is hence a more reliable indicator of a true overdensity.

Compared to the density, the velocity CPDFs shown in the right panels of \cref{fig:mock_pdf} are less skewed, displaying no noteworthy mismatch between conditional mean $\langle\vec{v}^\mathrm{true}\,|\,\hat{\vec{v}}\rangle$ and CPDF maximum. Accordingly, the range of likely true values around a given reconstructed velocity is approximately symmetric for both the NN and the WF. It is, however, significantly tighter for the NN, in particular for large reconstructed velocities (bottom-right panel).

\begin{figure}
    \centering
    \includegraphics[height=0.47\textwidth]{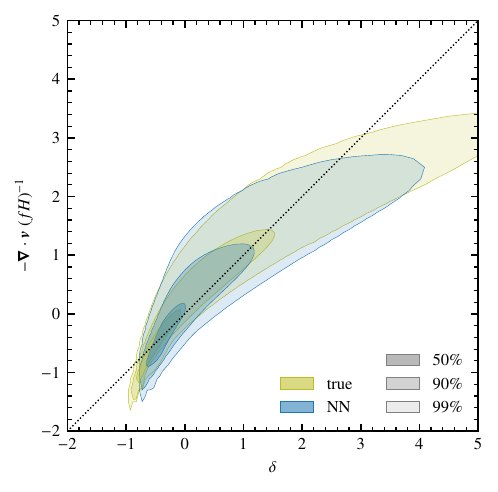}
    \caption{Distributions of velocity divergence versus density contrast values for the true (green) and NN-reconstructed (blue) fields in the validation mocks. They are represented as shaded areas bounded by the iso-density contours of the distribution. Going from a darker to lighter hue, the shaded areas contain $50\%$, $90\%$, and $99\%$ of all the points, respectively. The velocity divergence values are rescaled such that the diagonal dotted line marks the linear theory prediction.}
\label{fig:mock_velocity_vs_density}
\end{figure}

\begin{figure}
    \centering
    \includegraphics[height=0.47\textwidth]{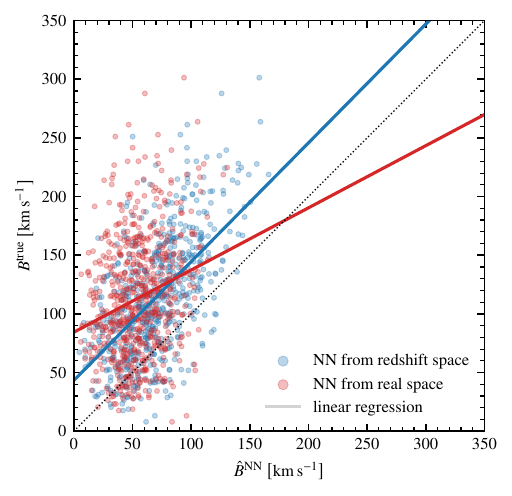}
    \caption{Scatter of true versus NN-reconstructed bulk flow amplitudes of the total survey volume using reconstructions from observations in redshift space (blue) and real space (red) in the validation mocks. The solid lines show the linear regression of true on reconstructed values, with slopes of $1.01$ and $0.53$ for redshift-space and real-space observations, respectively. For reference, the diagonal is marked with a dotted line.}
    \label{fig:mock_bulk_flow}
\end{figure}

\subsection{Velocity-density relation}
\label{sec:validation:velocity_vs_density}
A major advantage of the NN-reconstruction approach is that it does not assume any approximate (analytic) relation between the density and velocity fields. Instead, it learns this relation from the mocks, which in our case incorporate full nonlinear DM dynamics. On large scales $\gtrsim 15 \,h^{-1}\,\mathrm{Mpc}$, this relation is expected to be linear,
\begin{equation}
    \delta = \frac{-\vec{\nabla} \cdot \vec{v}}{fH} \,.
    \label{eq:linear_continuity_equation}
\end{equation}
where $H$ is the Hubble constant and $f = d \ln{D_+}/d \ln{a}$ is the growth rate defined as the logarithmic derivative of the linear growth factor $D_+$ with respect to the scale factor $a$ \citep{peebles_large-scale_1980}. Here, we adopted the expression $f = \Omega_\mathrm{m}^{0.55}$, which is a good approximation for $\Lambda$CDM cosmologies \citep{linder_cosmic_2005}. However, on the scale of $3 \,h^{-1}\,\mathrm{Mpc}$ probed by our reconstructions, the true velocity-density relation is nonlinear.

We tested if the NN reconstruction is in fact able to capture those nonlinearities by performing a point-by-point comparison between the velocity divergence and the density contrast. In \cref{fig:mock_velocity_vs_density}, we plot the iso-density contours of the resulting distribution that encompass 50, 90 and $99\%$ of all points. The velocity divergence was rescaled such that the linear theory relation \cref{eq:linear_continuity_equation} corresponds to the diagonal dotted black one-to-one line.

The true velocity-density distribution is shown in green, the NN-reconstructed one in blue. It is evident that both indeed exhibit substantial deviations from the linear theory relation, particularly for the high- and low-density tails of the distribution. Furthermore, we see that the NN reconstruction displays a remarkable qualitative agreement with the true distribution, verifying that the NN is able to capture the nonlinear small-scale density-velocity relation. Some quantitative discrepancies are expected, as the NN reconstructs the conditional mean of the true densities and velocities. We note that we refrained from plotting the velocity-density distribution of the WF, which, by construction, tightly scatters around the one-to-one linear theory line (the scatter only resulting from small numerical inaccuracies), to showcase the nonlinear true and NN-reconstructed distributions as clearly as possible.

\subsection{Probing ``super-survey'' scales}
\label{sec:validation:super_survey}
The density field derived from redshift surveys encodes information about ``super-survey'' mass fluctuations external to the observational volume \citep{nusserfeix2013,2018JCAP...02..022L,2018PhRvD..97f3527A,2020JCAP...10..007C}. This is because a galaxy's redshift incorporates its peculiar velocity, which is influenced by the cumulative gravity of the underlying density field both inside and outside the survey volume.

In particular, the bulk flow (mean velocity) of the whole survey volume is influenced by two factors: the gravitational pull of the external mass distribution and a contribution that is proportional to the radius vector of the center of mass inside the survey \citep{RomanJbullk}. The latter term can essentially be interpreted as the net gravitational force exerted on the mass fluctuations inside the survey by the homogeneous background density. Since the density in redshift space is affected by the bulk flow of the entire survey, the NN trained using the full velocity field as target should, at least partially, yield a velocity field that incorporates some constraints on the bulk flow.

\Cref{fig:mock_bulk_flow} presents a scatter plot of the true bulk flow amplitudes $B^\mathrm{true}$ versus NN-reconstructed bulk flow amplitudes $\hat{B}^\mathrm{NN}$, computed from the 640 mock galaxy catalogs of the validation set. The blue points are obtained from our usual NN-reconstructed velocity fields, where the NN was trained on the mock redshift-space observations described in \cref{sec:data:mock}. The red points are derived from velocity fields reconstructed by an NN that was trained on mock observations where the distribution of galaxies is given in real space (and no ZoA is imposed).

A visual inspection of the blue and red points clearly indicates that $B^\mathrm{true}$ is better correlated with $\hat{B}^\mathrm{NN}$ obtained from the distribution of galaxies in redshift space rather than real space. This is confirmed by the Pearson correlation coefficient of $0.59$ for the blue points, indicating a strong correlation between $B^\mathrm{true}$ and $\hat{B}^\mathrm{NN}$ in redshift space. Conversely, the Pearson coefficient for the red points is $0.24$, indicative of a weak correlation.

The blue and red straight lines in the figure correspond to linear regressions of $B^\mathrm{true}$ on $\hat{B}^\mathrm{NN}$. The slope of the blue line, obtained from the blue points, is $1.01$, indicating a close match between true and reconstructed values in redshift space. Furthermore, the slope of the red line is $0.53$, suggesting a less accurate reconstruction for galaxies in real space.

The results strongly indicate that the velocity field computed from NN reconstructions in redshift surveys indeed contains constraints on the bulk flow and hence external super-survey fluctuations. The weak correlation seen for reconstruction in real space is mainly the result of the term proportional to the radius vector of the center of mass inside the survey. Additionally, coherence of the density field may also contribute to the correlation seen for reconstructions in real space. Nonetheless, these effects are small compared to the correlation found for the redshift-space reconstruction.

\section{Reconstruction from the 2MRS data}
\label{sec_2mrs}
After having extensively validated the quality of the NN reconstruction using the mock data in \cref{sec:validation}, we are now applying the NN to the 2MRS dataset described in \cref{sec:data:2MRS}. In the following, we analyze the resulting density and peculiar velocity fields, reconstructed within a distance of $200 \,h^{-1}\,\mathrm{Mpc}$.\footnote{\label{note:public_fields}The reconstructed 3D fields and their uncertainties, estimated as the RMSE in the validation set of mocks, are publicly available at \href{https://github.com/rlilow/2MRS-NeuralNet}{https://github.com/rlilow/2MRS-NeuralNet}.} As for the mocks, the reconstructed 2MRS fields were all additionally smoothed with a Gaussian window of width $r_\mathrm{smooth} = 3 \,h^{-1}\,\mathrm{Mpc}.$

\begin{figure*}
    \centering
    \includegraphics[width=0.93\textwidth]{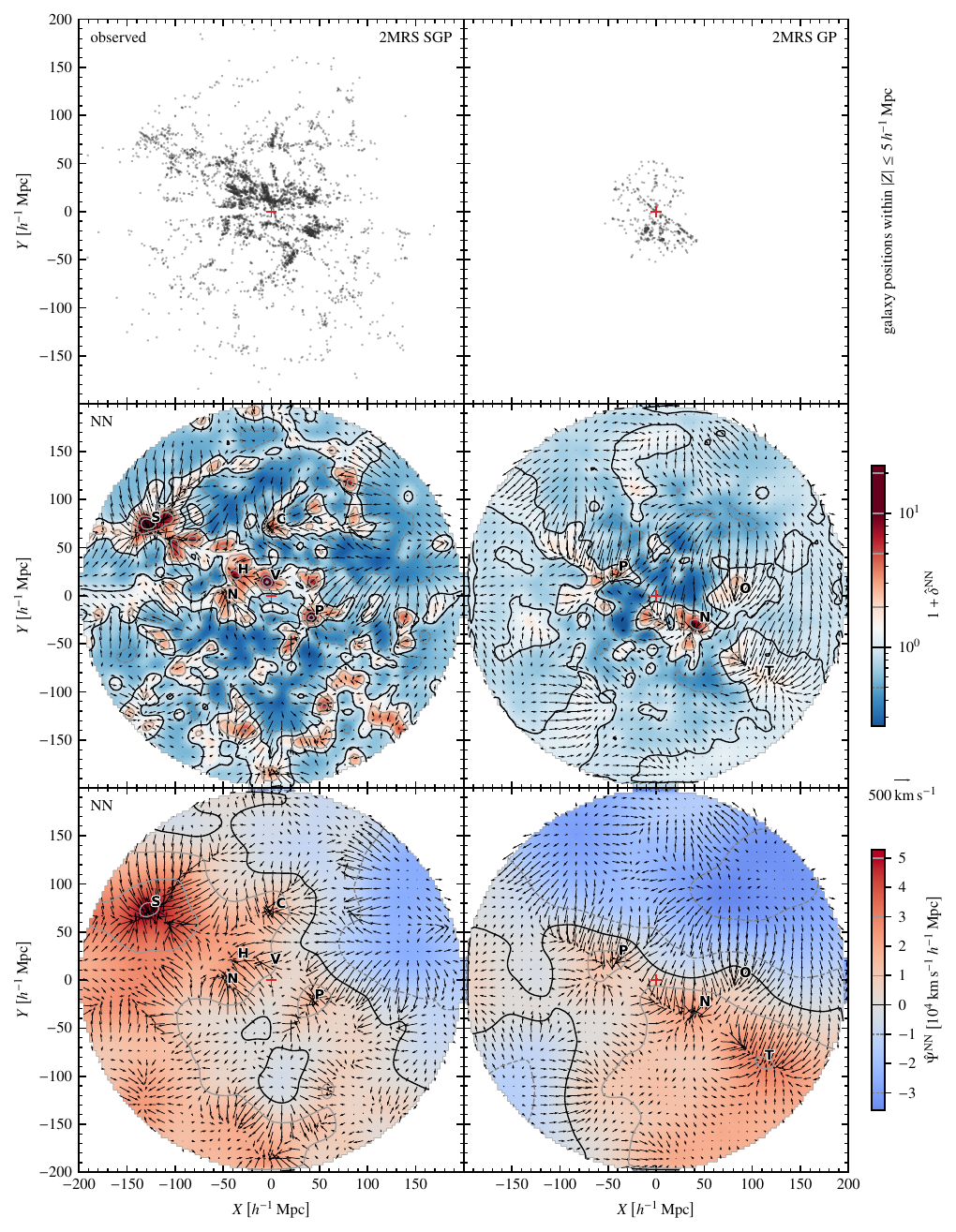}
    \caption{Observed 2MRS galaxies and reconstructed fields in a slice through the Supergalactic (left) and Galactic (right) planes. \textit{Top row:} redshift-space positions of the observed 2MRS galaxies within a distance of $5 \,h^{-1}\,\mathrm{Mpc}$ to the respective planes. The red cross marks the position of the observer (Local Group) at the origin. \textit{Middle row:} neural network-reconstructed density (heat map and contours) and peculiar velocity (arrows) fields. The values represented by the contours are marked in the color bar. A reference arrow representing a velocity of $500 \,\mathrm{km}\,\mathrm{s}^{-1}$ is shown on the right. Some dominant structures associated with known clusters are labeled by bold black letters: \textbf{S}hapley, \textbf{C}oma, \textbf{H}ydra-Centaurus, \textbf{V}irgo, \textbf{N}orma, \textbf{P}erseus-Pisces, \textbf{O}phiuchus, \textbf{T}riangulum Australis. \textit{Bottom row:} NN-reconstructed velocity potential field (heat map and contours) overlaid with the same peculiar velocity field shown in the middle row.}
    \label{fig:2MRS_field_slices}
\end{figure*}

\begin{table*}
    \caption{Positions and peculiar velocities of the NN-reconstructed density peaks that can be identified with known clusters.}
    \label{tab:clusters}
    \centering
    \begin{tabular}{lccccccc}
        \toprule
        Name & $\hat{r}^\mathrm{NN}$ & $\hat{l}^\mathrm{NN}$ & $\hat{b}^\mathrm{NN}$ & $\hat{v}_x^\mathrm{NN}$ & $\hat{v}_y^\mathrm{NN}$ & $\hat{v}_z^\mathrm{NN}$ & $\hat{v}_r^\mathrm{NN}$ \\
         & $[h^{-1}\,\mathrm{Mpc}]$ & $[{}^\circ]$ & $[{}^\circ]$ & $[\mathrm{km}\,\mathrm{s}^{-1}]$ & $[\mathrm{km}\,\mathrm{s}^{-1}]$ & $[\mathrm{km}\,\mathrm{s}^{-1}]$ & $[\mathrm{km}\,\mathrm{s}^{-1}]$ \\
        \midrule
        Shapley Supercluster (\object{SCl 124}) & $148$ & $312.0$ & $29.8$ & $-97 \pm 116$ & $-176 \pm 114$ & $-67 \pm 109$ & $24 \pm 121$ \\
        Coma Cluster (\object{ACO 1656}) & $73$ & $45.0$ & $88.3$ & $-107 \pm 72$ & $-169 \pm 64$ & $-83 \pm 61$ & $-90 \pm 61$ \\
        Centaurus Cluster (\object{ACO 3526}) & $35$ & $302.9$ & $24.0$ & $169 \pm 78$ & $-155 \pm 66$ & $89 \pm 67$ & $243 \pm 62$ \\
        Antlia Cluster (\object{ACO S636}) & $32$ & $273.0$ & $20.2$ & $267 \pm 82$ & $-39 \pm 66$ & $171 \pm 69$ & $104 \pm 62$ \\
        Norma Cluster (\object{ACO 3627}) & $52$ & $324.9$ & $-5.2$ & $-107 \pm 96$ & $-83 \pm 86$ & $211 \pm 68$ & $-63 \pm 85$ \\
        \object{Virgo Cluster} & $15$ & $288.4$ & $70.6$ & $18 \pm 90$ & $-525 \pm 79$ & $132 \pm 69$ & $272 \pm 69$ \\
        Perseus Cluster (\object{ACO 426}) & $49$ & $150.9$ & $-12.8$ & $-142 \pm 76$ & $-27 \pm 73$ & $-354 \pm 71$ & $185 \pm 70$ \\
        \object{Ophiuchus Cluster} & $87$ & $1.0$ & $9.3$ & $-55 \pm 118$ & $-266 \pm 105$ & $62 \pm 115$ & $-51 \pm 116$ \\
        \object{Triangulum Australis Cluster} & $151$ & $324.5$ & $-12.5$ & $-107 \pm 135$ & $-43 \pm 122$ & $456 \pm 123$ & $-157 \pm 135$ \\
        \bottomrule
    \end{tabular}
    \tablefoot{Galactic distances, longitudes, and latitudes of the reconstructed density peak positions as well as the Galactic Cartesian and radial  components of the reconstructed peculiar velocities relative to the CMB evaluated at these peak positions.}
\end{table*}

\subsection{Cosmography}
\label{sec:2mrs:cosmography}
In \cref{fig:2MRS_field_slices}, we present the reconstructed 2MRS density and velocity maps, both in the SGP (left panels) and the GP (right panels). For reference, the upper panels show the projected observed galaxy positions in redshift space in a slice extending $5\,h^{-1}\,\mathrm{Mpc}$ above and below the plane. In the SGP, containing a majority of the dominant structures in the local Universe, a rich distribution of galaxies can be seen near the origin that then rapidly thins out with increasing distance as the fraction of observable galaxies decreases. In the GP, which slices right through the ZoA, only a sparse sample of galaxies is observed near the origin where the ZoA is thinnest.

Maps of the NN-reconstructed density field are plotted in the middle panels of \cref{fig:2MRS_field_slices}. In the SGP, these maps reveal a vivid contrast of high-density regions, where known clusters reside, as well as filaments and voids. We label some of the dominant clusters that have been successfully reconstructed, listing their names in the figure caption. In the GP, the reconstructed density map is expectedly less rich in structures. As we have seen for the mock GP in \cref{fig:mock_field_slices}, only the most dominant large-scale structures are expected to be reconstructed, especially in the outer regions where the ZoA becomes wider. Nonetheless, a few known clusters can be identified, two of which, Norma and Perseus-Pisces, are also visible in the SGP. Two additional clusters, Ophiuchus and Triangulum Australis, can be identified with weaker but still pronounced overdensities, the latter at an remarkable distance of about $150 \,h^{-1}\,\mathrm{Mpc}$ deep into the ZoA.

The reconstructed peculiar velocity field, projected onto the plane, is overlaid as arrows. The flow pattern showcases how matter is displaced from under-dense regions into the filamentary structures and then funneled into the dense superclusters. This is particularly prominent in the filament to the top-right of Shapley in the SGP and in the filament to the top-left of Perseus-Pisces in the GP. In general, voids are clearly associated with diverging flows, while the highest overdensities act as main attractors with strongly converging peculiar velocities.

The dominant attractors are most easily identified in the bottom panels of \cref{fig:2MRS_field_slices}, where we display the reconstructed velocity potential, again overlaid with the velocity field. For this plot, the potential has been offset by a constant such that its average over the whole reconstruction volume vanishes. This way, the sources and sinks of the potential flow clearly correspond to negative and positive potential values, respectively. In the SGP, by far the most dominant attractor is Shapley, displaying a pronounced infall of matter from all directions. But also Coma, Perseus-Pisces and Norma (which is near the center of the Great Attractor) show a notable velocity convergence. Perseus-Pisces and Norma are also highlighted by their convergent flows in the GP. In addition, Triangulum Australis stands out as a dominant attractor in the GP at larger distances from the origin.

For all clusters labeled in \cref{fig:2MRS_field_slices} (or their major sub-clusters), we list the coordinates of their NN-reconstructed density peak as well as the NN-reconstructed peculiar velocity at those coordinates in \cref{tab:clusters}.

\subsection{Bulk flow and Local Group velocity}
\label{sec:2mrs:bulk_flow}
\begin{figure}
    \centering
    \includegraphics[width=0.49\textwidth]{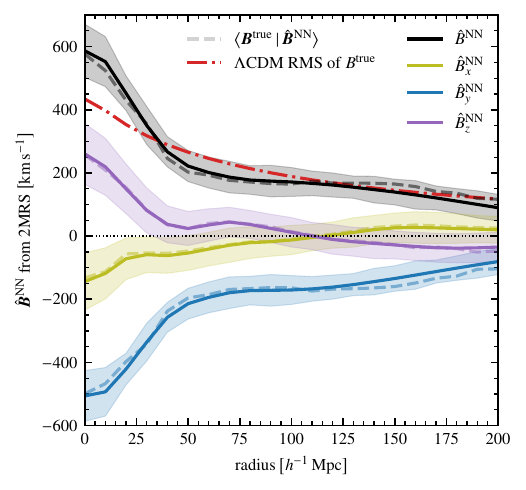}
    \caption{Neural network-reconstructed 2MRS bulk flow amplitude and components (black and colored solid lines, respectively) relative to the CMB in spheres of different radii around the Local Group. The shaded areas mark the $1\,\sigma$ uncertainty. The conditional mean of the true mock bulk flows given the reconstructed bulk flow is shown as dashed lines. The $\Lambda$CDM expectation of the true bulk flow RMS is shown as the red dash-dotted line. For reference, the horizontal dotted line marks a value of zero.}
    \label{fig:2MRS_bulk_flow}
\end{figure}

Given the significant correlation seen between NN-reconstructed and true bulk flows in the validation mocks in \cref{sec:validation:super_survey}, we are now investigating the bulk flow derived from the reconstructed 2MRS velocity field. \Cref{fig:2MRS_bulk_flow} plots the reconstructed bulk flow $\hat{\vec{B}}{}^\mathrm{NN}$ (solid lines) computed as the volume average of the velocity field $\hat{\vec{v}}^\mathrm{NN}$ in spheres of varying radius, showing its amplitude $\hat{B}^\mathrm{NN}$ and its three individual components $\hat{B}_x^\mathrm{NN}$, $\hat{B}_y^\mathrm{NN}$ and $\hat{B}_z^\mathrm{NN}$, in Galactic coordinates.

The reconstruction errors (shaded areas) were estimated as the standard deviation of the validation mock $\vec{B}^\mathrm{true}$ values conditioned on a small bin of mock $\hat{\vec{B}}{}^\mathrm{NN}$ values around those reconstructed from 2MRS. For a selection of radii, the reconstructed bulk flows and their errors are listed in \cref{tab:bulk_flows}.

We verify that the 2MRS-reconstructed bulk flow matches the conditional mean true bulk flow $\langle\vec{B}^\mathrm{true}\,|\,\hat{\vec{B}}{}^\mathrm{NN}\rangle$ (dashed lines) in good approximation, as expected for a mean posterior estimate. The minor deviations are a consequence of the finite sample size.

The bulk flow amplitude $\hat{B}^\mathrm{NN}$ in \cref{fig:2MRS_bulk_flow} first quickly drops from $\sim 590 \,\mathrm{km}\,\mathrm{s}^{-1}$ at radius $r=0$ to $\sim 220 \,\mathrm{km}\,\mathrm{s}^{-1}$ at $r=50 \,h^{-1}\,\mathrm{Mpc}$, then decreases slowly down to $\sim 90 \,\mathrm{km}\,\mathrm{s}^{-1}$ at $r=200\,h^{-1}\,\mathrm{Mpc}$. A similar behavior is seen in the individual Cartesian components, of which the Galactic $y$-component $\hat{B}_y^\mathrm{NN}$ has the dominant contribution. 

\Cref{fig:2MRS_bulk_flow} also shows, as a red dash-dotted line, the typical bulk flow amplitude expected in a $\Lambda$CDM cosmology with the cosmological parameters used in the Quijote simulations and assumed throughout this paper (cf.~\cref{sec:data:mock}). This expectation was computed as the linear-theory RMS velocity $\sigma_{\vec{v}}(r)$ in a sphere of radius $r$ via
\begin{equation}
    \sigma_{\vec{v}}^2(r) = \frac{H^2 \Omega_\mathrm{m}^{1.1}}{2 \uppi^2} \int_0^\infty \!\!\! \mathrm{d} k \, P_\delta(k) \, W_\mathrm{th}^2(k,r) \, W_\mathrm{g}^2(k) \,.
\end{equation}
Here, $P_\delta(k)$ is the matter density contrast power spectrum provided by the Cosmic Emu emulator \citep{moran_mira-titan_2023}, $W_\mathrm{th}(k,r)$ is a top-hat window function of radius $r$, and $W_\mathrm{g}(k)$ is a Gaussian window function of width $3 \,h^{-1}\,\mathrm{Mpc}$, accounting for the field smoothing. Given the uncertainty, the reconstructed bulk flow amplitude is found to be consistent with the $\Lambda$CDM expectation within $1\,\sigma$ for $r\gtrsim 25 \,h^{-1}\,\mathrm{Mpc}$ and within $2\,\sigma$ for smaller radii.

The bulk flow at $r=0$ (first line in \cref{tab:bulk_flows}) is simply the reconstructed peculiar velocity with respect to the CMB at the position of the observer, smoothed with a $3 \,h^{-1}\,\mathrm{Mpc}$ wide Gaussian window. It thus provides an estimate of the peculiar velocity of the Local Group. In both amplitude and latitude it is in close agreement with the observed Local Group velocity of $v_\mathrm{LG} = 620 \pm 15 \,\mathrm{km}\,\mathrm{s}^{-1}$ toward a Galactic longitude and latitude of $l_\mathrm{LG} = 271.9 \pm 2.0^\circ$ and $b_\mathrm{LG} = 29.6 \pm 1.4^\circ$ \citep{aghanim_planck_2020}. In longitude it is off by $\sim 18^\circ$, which is a $1.8\,\sigma$ deviation.

\begin{table*}
    \caption{Neural network-reconstructed bulk flows in spheres of different radii around the Local Group.}
    \label{tab:bulk_flows}
    \centering
    \begin{tabular}{ccccccc}
        \toprule
        Radius & $\hat{B}_x^\mathrm{NN}$ & $\hat{B}_y^\mathrm{NN}$ & $\hat{B}_z^\mathrm{NN}$ & $\hat{B}^\mathrm{NN}$ & $\hat{l}^\mathrm{NN}$ & $\hat{b}^\mathrm{NN}$ \\
        $[h^{-1}\,\mathrm{Mpc}]$ & $[\mathrm{km}\,\mathrm{s}^{-1}]$ & $[\mathrm{km}\,\mathrm{s}^{-1}]$ & $[\mathrm{km}\,\mathrm{s}^{-1}]$ & $[\mathrm{km}\,\mathrm{s}^{-1}]$ & $[{}^\circ]$ & $[{}^\circ]$ \\
        \midrule
        0 & $-144 \pm 91$ & $-506 \pm 80$ & $260 \pm 96$ & $587 \pm 84$ & $254.1 \pm 10.0$ & $26.3 \pm 9.3$ \\
        10 & $-118 \pm 81$ & $-493 \pm 77$ & $219 \pm 90$ & $552 \pm 79$ & $256.5 \pm 9.6$ & $23.4 \pm 9.6$ \\
        20 & $-72 \pm 66$ & $-421 \pm 50$ & $150 \pm 78$ & $452 \pm 55$ & $260.3 \pm 9.4$ & $19.3 \pm 10.1$ \\
        30 & $-59 \pm 57$ & $-335 \pm 60$ & $81 \pm 71$ & $350 \pm 60$ & $260.1 \pm 9.4$ & $13.5 \pm 11.4$ \\
        40 & $-62 \pm 56$ & $-257 \pm 47$ & $37 \pm 58$ & $267 \pm 48$ & $256.5 \pm 13.0$ & $8.0 \pm 13.6$ \\
        50 & $-54 \pm 55$ & $-214 \pm 49$ & $24 \pm 54$ & $222 \pm 49$ & $255.9 \pm 15.5$ & $6.3 \pm 15.2$ \\
        100 & $-12 \pm 46$ & $-171 \pm 46$ & $12 \pm 47$ & $171 \pm 46$ & $265.9 \pm 16.2$ & $4.1 \pm 16.2$ \\
        150 & $26 \pm 47$ & $-134 \pm 40$ & $-28 \pm 44$ & $140 \pm 41$ & $281.1 \pm 16.4$ & $-11.5 \pm 15.3$ \\
        200 & $20 \pm 42$ & $-81 \pm 41$ & $-35 \pm 47$ & $90 \pm 42$ & $284.0 \pm 22.7$ & $-22.6 \pm 22.6$ \\
        \bottomrule
    \end{tabular}
    \tablefoot{Cartesian Galactic components, amplitude, and direction in Galactic longitude and latitude of the reconstructed bulk flows relative to the CMB.}
\end{table*}

\section{Summary and discussions}
\label{sec:conclusion}
In \citetalias{ganeshaiah_veena_large-scale_2023}, preliminary tests were conducted to explore the reconstruction of large-scale structure density and velocity fields using an NN framework. These initial tests involved training and validation data that were derived using approximate gravitational dynamics, which evolved from $\Lambda$CDM initial conditions using second-order Lagrangian perturbation theory. Only a simplified RSD model was employed, and galaxy biasing and selection effects were not taken into account. The primary objectives of \citetalias{ganeshaiah_veena_large-scale_2023} were to assess the feasibility of using NNs for modelling the mapping between observations and underlying matter fields, albeit in an approximate form, and crucially, to elucidate the relationship between the reconstructed fields and well-known statistical estimates, namely the mean posterior estimate and the WF.

The current study significantly builds upon \citetalias{ganeshaiah_veena_large-scale_2023}, aiming to develop an NN framework that can be applied to actual observations of the distribution of galaxies in redshift space. This effort involves generating mock catalogs that closely mirror the characteristics of realistic redshift surveys. Specifically, we meticulously created mock catalogs that capture the key features of 2MRS. These catalogs, which are essential for NN training and validation, were derived from the full high-resolution N-body Quijote simulations \citep{villaescusa-navarro_quijote_2020}. They include full RSDs, a nonlinear galaxy bias model, a radially declining selection function, and masking by the ZoA, all calibrated to mimic the properties of the actual 2MRS as closely as possible.

\subsection{Method}
Our NN is an autoencoder with U-Net architecture, designed to discern the nonlinear mappings between the input observed galaxy number count field and the target true matter density and velocity fields. To achieve this, two separate NNs were trained, each minimizing a customized loss functions designed for predicting either the density or velocity fields. Both loss functions compute a weighted mean squared error between the true and reconstructed fields. Incorporating a distance-dependent weighting into our loss function maintains the property that during training the NN reconstructions converge toward the mean posterior estimate of the true fields given an observation. Our methodology hence results in a robust nonlinear statistical estimator that is computationally expensive to implement with non-machine learning techniques.

We found that customizing the weighting significantly enhanced the NN performance. It accelerated the training convergence and reduced the achieved reconstruction error. The empirically determined weighting is proportional to the selection function, thus accounting for the radially increasing shot noise, and additionally scales with the inverse distance for the velocity reconstruction. Moreover, by assuming potential flow in the loss function for predicting peculiar velocities, we are eliminating the possibility of predicting any unphysical rotational component in the reconstructed velocity field.

\subsection{Validation}
Prior to applying our trained models to the 2MRS data, we conducted a comparison with standard WF reconstructions on a validation set of our Quijote-based mocks. This comparison serves as a benchmark for our model, allowing us to discern the advantages and potential limitations of applying an NN instead of a traditional WF. 

It was shown in \citetalias{ganeshaiah_veena_large-scale_2023} that the NN consistently outperforms WF in terms of reconstruction accuracy. Similarly, in this study, we find that throughout the survey volume the fields reconstructed by the NN exhibit a lower RMSE compared to those reconstructed by the WF, while also matching the true density and velocity RMS values more closely (cf.~\cref{fig:mock_rms_rmse}). Specifically, the RMSE of the NN-reconstructed density increases from $\sim 0.5$ near the observer to $\sim 0.85$ at a distance of $200 \,h^{-1}\,\mathrm{Mpc}$ (compared to a true RMS value of $\sim 1.05$). Over the same distance range the RMSE of the NN-reconstructed velocity components increases from $\sim 80 \,\mathrm{km}\,\mathrm{s}^{-1}$ to $\sim 200 \,\mathrm{km}\,\mathrm{s}^{-1}$ (compared to a true RMS value of $\sim 270 \,\mathrm{km}\,\mathrm{s}^{-1}$).

By means of a detailed point-by-point comparison we also demonstrated that the distribution of true versus reconstructed field values is consistently tighter for the NN than for the WF (cf.~\cref{fig:mock_true_vs_reconstructed,fig:mock_pdf}). Furthermore, we verified in this comparison that the conditional mean of true field values for a given reconstructed value closely matches that reconstructed value. In other words, the line of linear regression of true on reconstructed field values closely follows the identity line. This demonstrates that the NN reconstruction shows the behavior expected for a mean posterior estimate.

We demonstrated by visual inspection that the NN-reconstructed fields successfully recover the true large-scale structures in the validation mocks, capturing high-density regions, filamentary features and voids, as well as their corresponding converging and diverging velocity flows (cf.~\cref{fig:mock_field_slices}).

To explicitly test the capability of the NN in capturing nonlinear effects, we analyzed the distribution of velocity divergence versus density contrast (cf.~\cref{fig:mock_velocity_vs_density}). We found a good agreement between the true and reconstructed distributions, with both displaying a clear deviation from the linear theory relation.

\subsection{Super-survey scales}
A galaxy's peculiar velocity is influenced by the gravitational pull of mass fluctuations outside the survey volume. While the distribution of galaxies in real space remains unaffected by external super-survey scales, the redshift-space density field encodes signatures of these scales through the dependence of galaxy redshifts on peculiar motions.

We showed that our approach to training the NN produces velocity fields that indeed contain information about super-survey scales. This was demonstrated using the validation mock data, through the significant correlation between the bulk flow computed from the reconstructed velocity and the true bulk flow (cf.~\cref{fig:mock_bulk_flow}). This correlation was found only for reconstructions from the distribution of galaxies in redshift space, whereas for reconstructions from real space the correlation was weak. In future work, we intend to train a dedicated NN to predict bulk flows in order to probe super-survey scales more effectively.

\subsection{2MRS application}
The extensively trained and validated NN was applied to 2MRS, to reconstruct the 3D density and peculiar velocity fields in the local Universe out to a distance of $200 \,h^{-1}\,\mathrm{Mpc}$. The resulting fields, defined on a regular cubic $128\times128\times128$ grid of side length $400 \,h^{-1}\,\mathrm{Mpc}$, as well as their per-grid-point uncertainties, estimated as the RMSE in the mock validation set, are publicly available (cf.~\cref{note:public_fields}).

The reconstructed fields recover multiple known clusters, mostly in the SGP, where Shapley is found to be the dominant attractor (cf.~\cref{fig:2MRS_field_slices}). But we also see a few clusters in the GP, which cuts directly through the ZoA. Most notably, despite residing at the edge of the ZoA, Triangulum Australis at a distance of $\sim 150 \,h^{-1}\,\mathrm{Mpc}$ can be identified by a pronounced convergence in the velocity field as well as a minor overdensity in the reconstruction.

From the reconstructed velocity field, we furthermore computed the bulk flow within spheres of varying radii $r$ around the observer (cf.~\cref{fig:2MRS_bulk_flow}). It shows a remarkable similarity to the bulk flow reconstructed previously in \citet[hereafter \citetalias{lilow_constrained_2021}]{lilow_constrained_2021}. \citetalias{lilow_constrained_2021} combined a WF-reconstructed velocity field from 2MRS with an external bulk flow component calibrated by matching the observed peculiar velocities from the galaxy distance catalog Cosmicflows-3 \citep{tully_cosmicflows-3_2016}. In addition, the comparison with Cosmicflows-3 velocities was used to optimize the value of the normalized growth rate $f \sigma_8$, which determines the peculiar velocity amplitude.

For $r \lesssim 100 \,h^{-1}\,\mathrm{Mpc}$, the two bulk flow reconstructions are found to be in 1 to $2\,\sigma$ agreement. For $r \gtrsim 100 \,h^{-1}\,\mathrm{Mpc}$, the NN-reconstructed bulk flow amplitude drops more severely than the reconstruction obtained in \citetalias{lilow_constrained_2021}. This could be the consequence of an only partially captured 2MRS-external bulk flow contribution by the NN. Alternatively or additionally, it might be due to the approximation of the external velocity field contribution in \citetalias{lilow_constrained_2021} as a constant (dipole) bulk flow term, which is a better approximation near the center of the reconstruction volume than toward its boundary.

Furthermore, the NN-reconstructed peculiar velocity of the Local Group, corresponding to the bulk flow at $r=0$, was found to be $\hat{v}^\mathrm{NN}_\mathrm{LG} = 587 \pm 84 \,\mathrm{km}\,\mathrm{s}^{-1}$ toward a Galactic longitude and latitude of $\hat{l}^\mathrm{NN}_\mathrm{LG} = 254 \pm 10^\circ$ and $\hat{b}^\mathrm{NN}_\mathrm{LG} = 26 \pm 9^\circ$, respectively. This closely matches the observed Local Group velocity of $v_\mathrm{LG} = 620 \pm 15 \,\mathrm{km}\,\mathrm{s}^{-1}$ toward $l_\mathrm{LG} = 271.9 \pm 2.0^\circ$ and $b_\mathrm{LG} = 29.6 \pm 1.4^\circ$ \citep{aghanim_planck_2020}, apart from an $18^\circ$ deviation in longitude ($1.8\,\sigma$).

These results are remarkable since our NN reconstruction only uses the 2MRS galaxy redshifts as input. All information about 2MRS-external bulk flow contributions must thus be inferred from their imprint on the observed RSDs.

We need to point out, however, that the reconstructions are dependent on the prior assumptions entering our training mocks, in particular assuming a fixed fiducial $\Lambda$CDM cosmology (cf.~\cref{sec:data:mock}). This fixes the value of the normalized growth rate to $f \sigma_8 =0.444$. Training the NN on mocks assuming other values of $f \sigma_8$ would likely affect at least the amplitude of the reconstructed velocity field. In subsequent work, this could be studied by employing mocks with varying cosmological parameters, for example the latin-hypercube set of Quijote \citep{villaescusa-navarro_quijote_2020}.

\subsection{Conclusion}
There are evident advantages to using an NN for reconstructions over standard methods. First, the training process does not rely on any approximations for the gravitational dynamics. By incorporating a sufficient number of training data and model parameters, the NN can effectively learn to perform reconstructions by capturing key features of full N-body dynamics, including modified gravity models. In our application, this is clearly demonstrated in \cref{fig:mock_velocity_vs_density}, showing the nonlinear relation between the NN-reconstructed velocity and density. This further implies that the NN is capable of probing intermediate (transitional) scales between linear large scales and highly nonlinear small scales. Second, incorporating galaxy bias, RSDs, and selection characteristics of the survey is relatively straightforward in mock catalogs. Consequently, the NN can learn how to extract information while accounting for the limitations of the actual data. Third, by using the appropriate loss and activation functions, an NN can avoid unphysical predictions, such as negative densities or rotational velocity components. These choices can be adapted to any problem and contribute to the NN's ability to generate consistent and physically relevant results. 

In contrast, standard methods often require explicit assumptions about all of these ingredients, which NNs can easily tackle through well-constructed mocks. Given the complexity of these components, standard methods typically resort to simplistic assumptions, which may introduce statistical biases into the inferred information.

Nonetheless, it is crucial to acknowledge that the NN methodology is not without its challenges. For example, acquiring a sufficiently large set of training data, especially when exploring a broad range of cosmological parameters, remains a difficult task. This is in contrast to standard methods, where explicit dependencies on these parameters are incorporated into the approximations that are used, such as the dependence on the background matter density parameter in the velocity-density relationship.

Similarly, accounting for all potentially relevant observational effects is challenging. In constructing the mock data, for example, we did not make a special effort to mimic the properties of the actual Local Group of galaxies. The flow of galaxies in the vicinity of the Local Group is remarkably quiet up to a distance of $5\,h^{-1}\,\mathrm{Mpc}$, exhibiting a coherent velocity $v_\mathrm{LG} \approx 600 \,\mathrm{km}\,\mathrm{s}^{-1}$ with respect to the CMB. This coherent flow introduces a strong artificial dipole density enhancement in the distribution of galaxies in redshift space, with redshifts measured in the CMB frame. It is possible to define redshifts with respect to the Local Group motion, which would yield a redshift-space galaxy distribution free from a local dipole enhancement. However, it turns out that it is not possible to extract from the Quijote simulation suite a sufficiently large number of training mocks matching the observed local environment in terms of flow coherence and moderate density. Therefore, we opted to stick to using CMB redshifts since a well-trained NN should in principle be able to learn how to disentangle the artificial local density dipole. 

Overall, NN methods offer a wide range of possibilities and enhancements in the field of LSS reconstructions. For example, NN-reconstructed density and velocity maps of the local Universe can improve the extraction of nonlinear cosmographic features, such as filaments, walls, and clusters. The improved reconstructed velocity fields can also be used to test the validity of different gravity models by performing standard comparisons with the observed velocity inferred from distance measurements \citep[e.g.][]{davis_local_2011,turnbull_cosmic_2012,ma_comparison_2012,carrick_cosmological_2015,said_joint_2020,boruah_cosmic_2020,stahl_peculiar-velocity_2021,lilow_constrained_2021,hollinger_cosmological_2024}. Furthermore, NN-reconstructed nonlinear peculiar velocities can be employed for peculiar velocity corrections in local measurements of the Hubble constant using supernovae or other distance measurements \citep[e.g.][]{peterson_pantheon_2022,riess_comprehensive_2022,kenworthy_measurements_2022,brout_pantheon_2022}. In the future, we plan to extend the applicability of the presented method by training on mocks with a wider range of cosmologies as well as incorporating the observational characteristics and systematics of other surveys.

\begin{acknowledgements}
    P.G.V.~is supported in part at the Technion by a fellowship from the Lady Davis Foundation. We thank Francisco Villaescusa Navarro and Shy Genel for help with the Quijote simulations. This research is supported by a grant from the Israeli Science Foundation and a grant from the Asher Space Research Institute.
\end{acknowledgements}

\bibliographystyle{aa}
\bibliography{bibliography_PGV,bibliography_RL,bibliography_AN}

\end{document}